\documentstyle[12pt, aasms4, epsfig]{article}
\catcode`@=11
\def\eqalign#1{\null\,\vcenter{\openup\jot\m@th
  \ialign{\strut\hfil$\displaystyle{##}$&$\displaystyle{{}##}$\hfil
      \crcr#1\crcr}}\,}
\def\eqalignleft#1{\null\,\vcenter{\openup\jot\m@th
  \ialign{\strut$\displaystyle{##}$\hfil&$\displaystyle{{}##}$\hfil
      \crcr#1\crcr}}\,}
\catcode`@=12

\def\lax    {\ifmmode{_<\atop^{\sim}}\else{${_<\atop^{\sim}}$}\fi}
\def\gax    {\ifmmode{_>\atop^{\sim}}\else{${_>\atop^{\sim}}$}\fi}
\def\kms    {\ifmmode{{\rm ~km~s}^{-1}}\else{~km~s$^{-1}$}\fi}

\def\bk{\lower 6pt\hbox{${\buildrel k\over \sim}$}}
\def\bv{\lower 6pt\hbox{${\buildrel v\over \sim}$}}
\def\ul#1{$\underline{\smash{\vphantom{y}\hbox{#1}}}$}

\begin{document}
\doublespace
\textwidth 6.7truein
\textheight 9.5truein
\topmargin -1cm
\hoffset = 0.5truein
\title{\bf
DISTRIBUTION OF DUST FROM KUIPER BELT OBJECTS
}
\author{\bf Nick N. Gorkavyi\altaffilmark{1}}
\affil{
Laboratory for Astronomy and Solar Physics, NASA/Goddard Space Flight
Center\\ Greenbelt, MD 20771
}
\altaffiltext{1}{NRC/NAS Senior Research Associate; e-mail:
gorkavyi@stars.gsfc.nasa.gov}
\author{\bf Leonid M. Ozernoy\altaffilmark{2}}
\affil{5C3, School of Computational Sciences and Department of Physics
\& Astronomy,\\ George Mason U., Fairfax, VA 22030-4444; also Laboratory for
Astronomy and Solar\\ Physics, NASA/Goddard Space Flight Center, Greenbelt,
MD 20771}
\altaffiltext{2}{Corresponding author. Fax: $+1$-301-286-1617; e-mail:
ozernoy@science.gmu.edu, ozernoy@stars.gsfc.nasa.gov}
\author{\bf Tanya Taidakova\altaffilmark{3}}
\affil{
Computational Consulting Service, College Park, MD 20740
}
\altaffiltext{3}{e-mail: simeiz@aol.com}
\author{\bf John C. Mather\altaffilmark{4}}
\affil{
Laboratory for Astronomy and Solar Physics, NASA/Goddard Space Flight
Center\\ Greenbelt, MD 20771}
\altaffiltext{4}{e-mail:
john.c.mather@gsfc.nasa.gov}
\bigskip
\newpage
\centerline{\bf Abstract}

Using an efficient computational approach, we have reconstructed the
structure of the dust cloud in the Solar system between 0.5 and 100 AU
produced by the Kuiper belt objects.
Our simulations offer a 3-D physical model of the `kuiperoidal' dust
cloud based on the distribution of  280 dust particle trajectories 
produced by 100 known Kuiper belt objects ; the resulting 3-D grid consists
of $1.9\times 10^6$ cells containing $1.2\times 10^{11}$ particle positions.
The following processes that influence the dust particle dynamics are
taken into account: 1) gravitational scattering on the eight planets 
(neglecting Pluto); 2) planetary resonances; 3) radiation pressure;  
and 4) the Poynting-Robertson (P-R) and solar wind drags.
We find the dust distribution highly non-uniform:
there is a minimum  in the kuiperoidal dust between Mars and Jupiter, 
after which both the column and number densities of kuiperoidal 
dust sharply increase with heliocentric distance between 5 and 10 AU,
and then form a plateau between 10 and 50 AU. Between 25 and 45 AU, 
there is an appreciable concentration of kuiperoidal dust in the form of 
a broad belt of mostly resonant particles associated with Neptune.
In fact, each giant planet possesses its own
circumsolar dust belt consisting of both resonant
and gravitationally scattered particles.
As with the cometary belts simulated in our related papers
(Ozernoy, Gorkavyi, \& Taidakova 2000a,b), we reveal a
rich and sophisticated resonant structure of
the dust belts containing families of resonant peaks and gaps.
An important result is that both the column and number dust density are more 
or less flat between 10 and 50 AU, which  might explain 
the surprising data obtained by Pioneers 10 \& 11 and Voyager that the dust 
number density remains approximately distance-independent in this region.
The simulated kuiperoidal dust, in addition to asteroidal and cometary dust,
might represent a third possible source of the zodiacal light in
the  Solar system.


\newpage
\section*{1. Introduction}
It is well known that the Edgeworth-Kuiper belt objects
(called `kuiperoids' hereinafter) can 
replenish the cometary populations throughout the Solar system
(e.g., Levison \& Duncan 1997; Ozernoy, Gorkavyi, \& Taidakova 2000a,b
$\equiv$ OGT 2000a,b; and refs. therein). Recently, it has been  recognized 
that kuiperoids might be also one of the major sources of dust in the Solar 
system (e.g., Backman et al. 1995, Liou et al. 1996).
This dust can be produced due to evaporation
of the volatile material from the surface of kuiperoids
by processes, such as the solar radiation and wind,
mutual collisions of kuiperoids, micrometeor bombardment, etc.
In this paper,  we perform extensive numerical simulations to examine the
distributions (both in orbital parameters and in space) of kuiperoidal
dust particles and thereby
to analyse the structure of the kuiperoidal dust cloud.

The dynamics of this dust is determined by three major effects:
(i)~the Poynting-Robertson (P-R) drag (including radiation pressure and
solar wind drag), (ii)~gravitational scattering on the planets, and 
(iii)~resonances with the planets. Extensive work on dust particle 
evolution governed by these effects has been done by a number of
investigators (Weidenschilling \& Jackson 1993; Hamilton 1994;
Roques et al. 1994; Liou \& Zook 1997; Gorkavyi, Ozernoy, Mather, \& Taidakova 
($\equiv$ GOMT) 1997a,b and 1998a,b; Kortenkamp \& Dermott 1998). The evolution
of kuiperoidal dust 
was analyzed by Liou, Zook,  \& Dermott  (1997), Gorkavyi, Ozernoy, \& 
Taidakova (1998), and  Liou \& Zook (1999). 
The present paper makes a next step 
by employing a fast and efficient method to
compute a stationary distribution function of dust particles in the phase
space, which provides much better statistics to derive a 3-D model of the 
interplanetary dust cloud and to reveal its rich resonant structure.

In Sect.~2, we discuss the sources of dust in the outer Solar system.
The dynamical evolution of dust particles is reviewed in Sect.~3. Sect.~4 
describes our numerical method that enables us to  compute the 
3-D distribution of dust in the Solar system. We employ an implicit 
second-order integrator for dissipative systems (Taidakova \& Gorkavyi 1999) 
outlined in Appendix. Sect.~5 contains
the results of these computations, which reveal the global dust distribution
as well as interesting details of its resonant structure. Our conclusions
are presented in Sect.~6.
\section*{2. The Sources of Dust Particles}

There is mounting evidence that the sources of the interplanetary dust
particles (IDPs) cannot be entirely reduced simply to
those comets which produce the observed dust tails and/or to asteroids
which are thought to be responsible for the observed `dust bands'
in the IDP emission. Many facts forces us to suspect
that additional sources of the interplanetary dust must exist.
Among others, two groups of facts are worth mentioning: (i)~According to 
Pioneers 10 \& 11 and Voyagers  1 \& 2 data, the dust number density 
is approximately distance-independent in the outer Solar system between 
10 and 40-50 AU (Humes 1980, Divine 1993, Gurnett et al. 1997), while both 
the asteroidal and cometary dust number densities are known to sharply 
decrease with heliocentric distance (GOMT 1997b); (ii)~Chemical analyses 
and other space-based data indicate
that some IDPs spent a much larger time  in space than the
typical asteroidal and cometary particles (Flynn 1996). Thus, there
is  strong evidence in favor of other sources of dust in the Solar system, 
along with the known comets and asteroids.

This third component
of the IDP cloud might be the `kuiperoidal' dust (Backman et al. 1995).
The Kuiper belt can influence the formation of the IDP cloud in two ways:
1) as a source of small-size particles slowly drifting toward the
Sun under the combined action of the PR-drag, gravitational scattering,
and resonances;
and 2) as a source of trans-Jovian comets. It is commonly agreed that
the Jupiter-family comets are produced by transporting the comets from
the Kuiper belt via gravitational scattering on the four giant planets so that
each planet scatters the comets both toward and away from the Sun (Levison
\& Duncan 1997). Our  simulations  (OGT 2000a,b) 
indicate that, between Jupiter and Neptune, there is
a large population of minor bodies forming four cometary-asteroidal
belts near the orbits of the giant planets.

According to our simulations, the minor body families of Saturn, Uranus, and 
Neptune should contain progressively  larger numbers of comets than one sees  near
Jupiter. Even despite a many-fold decrease of the solar heat intensity
at such large distances, those numerous comets may produce dust
in amounts comparable to that from a few active J-comets.
Complementary mechanisms of dust release from kuiperoids and Centaurs
between Jupiter and Neptune can include
impacts of large grains and the  solar wind. Without discussing here 
the dust production by the above mechanisms, which is out of
our paper's scope, we simply
refer to observational data that indicate,  for a number of kuiperoids and
Centaurs,  a steady cometary activity lasting for years
(e.g. Brown \& Luu 1998 and refs. therein).
\section*{3. Dynamical evolution of dust particles}
The kuiperoidal dust experiences the same dynamical effects as the
cometary and asteroidal dust, with the only difference
that, due to a slower PR-drift and a stronger influence of the giant
planets, the role of gravitational scattering and resonance captures
must be more important for it.

Just after the birth of a dust particle, the solar pressure forces
it to change its orbit to a more distant and eccentric one.
This change of the orbital parameters (semimajor axis $a_{d}$ and
eccentricity $e_{d}$) of dust particles, which start their journey
from the apocenter or pericenter of kuiperoidal orbits, is described by
$$
  a_{d}=a_K{1-\beta\over 1-2\beta\left(1\pm e_K\right)^{-1}}, \eqno (1)
$$
$$e_{d}=\left(1-{(1-e_K^2)\left [1-2\beta\left(1\pm e_K\right)^{-1}\right]
\over(1-\beta)^2}\right)^{1/2}.  \eqno   (2)
$$
Here $a_K$ and $e_K$ are the orbital parameters of kuiperoids,
signs ``$+$" and ``$-$" correspond to the start from apocenter and
pericenter, accordingly, and $\beta\approx L_\odot/(M_\odot r)$ is the
ratio of the solar light pressure and the gravitational force applied
to the dust grain of radius $r$ (in $\mu$m).
In what follows, we consider two  $\beta$-values: $\beta=0.285$ and
$\beta=0.057$, which correspond, for grains of density $\rho=(1-2)$ g/cm$^3$,
to $r =(1-2)~\mu$m and $r =(5-10)~\mu$m, respectively.

The basic mechanisms of the dynamical evolution of dust are as follows:

1. {\it Drift of particles toward the Sun under the Poynting-Robertson drag}. 
Such drift occurs more or less freely except for short periods when particles
experience a strong gravitational scattering on the planets or when they
pass through the outer or inner resonances without being captured into them.
A particle governed by the P-R drag drifts
in the $(a,e,i)$-space, with a decreasing $a$ and $e$,
along the line
$$             a(1-e^2)e^{-4/5}={\rm const}, ~~i={\rm const}. \eqno (3)
$$

    2. {\it Gravitational scattering by the planets}, which builds up a
population of particles with large eccentricities, including
    grains that are ejected from the Solar system.
A gravitationally scattered particle experiences a rather chaotic `jump' 
in the $(a,e,i)$-space, but conserves its Tisserand parameter ${\cal T}$
during the jump:
$${\cal T}={1\over 2}\left({a_{pl}\over a}\right)+\left({a\over 
a_{pl}}\right)^{1/2}\left(1-e^2\right)^{1/2}\cos i={\rm const.}  \eqno (4)$$
Although the P-R drag changes that parameter, it happens  on a much 
longer timescale.  As shown in  OGT (2000b), gravitational 
scattering results in the motion of most
of the particles within the so called `crossing zone'
(i.e. the zone of strong gravitational scattering) defined,
in the $(a,e)$-plane of orbital coordinates, by:
$$a(1-e)\leq a_{pl} ~~~{\rm if}~ a>a_{pl},$$
$$a(1+e)\geq a_{pl} ~~~{\rm if}~ a<a_{pl}.  \eqno (5) $$
Here $a$ and $e$ are the semimajor axis and eccentricity of the test 
particle, respectively; and
$a_{pl}$ is the semi-major axis of the planet.
As a result of gravitational scatterings, the particle pericenters are 
close enough to the planet's orbit $a_{pl}$ (OGT 2000b):
 $$ a (1-e) \approx  a_{pl} \quad {\rm if}~ a>a_{pl}.   \eqno (6)$$

3. {\it Resonant capture of dust into outer resonances with the planets}.
A particle  captured into a resonance is positioned on the line
     $$           a=(1-\beta)^{1/3}\left({n\over m}\right)^{2/3}a_{pl},
     \eqno (7)
$$
where usually $n>m$ (i.e. this is an outer  resonance). The
eccentricity of the resonant particle increases with time
(Weidenschilling and Jackson 1993), while its $i$ oscillates and gradually
decreases (Liou \& Zook 1997).

Athough the above dynamical trajectories governed by just one dominating 
dynamical factor turn out to be only approximate as soon as the two
other factors are taken into account, they are very helpful as analytical
approximations in the appropriate limiting cases. Moreover, in accordance
with the dominating dynamical factor, we expect to get
three major components of dust populations:  i)~`freely' drifting particles,
(ii)~gravitationally scattered particles, and (iii)~particles captured
into resonances. This classification is helpful while interpreting
the results of  the present numerical simulations.
In Sect.~5, we consider the steady-state distribution of these 
three dust components, but before that
we describe  our computational approach.

\section*{4. Computational Method: Simulation of a Quasi-stationary
Distribution of Dust Particles in the Solar System}

We calculate the orbital elements $(a,e,i)$ and then the spatial
positions of massless particles  starting from a
particular kuiperoid as the source of dust  and  drifting toward the Sun
under the P-R drag. On its way to the Sun, each particle 
undergoes the gravitational influence of the planets. To save
computational effort, we assume that the Sun is fixed at the origin and the 8 
planets (excluding Pluto) are on circular orbits with zero
inclinations (this approximation will be abandoned in our
further work).

The distribution of dust in the Solar system averaged over a time scale
of planetary orbital motions is described as stationary.
To simulate this 
stationary distribution of dust particles, we applied the following 
procedure: We computed the dynamical trajectory of each particle and kept 
a record 
of the particle's orbital elements and positions with 
a certain time interval. These data were then used to characterize the 
orbital elements and positions
of {\it many} particles over the entire time span, from
an initial instant intil the particle's death (impact on planet,
the Sun, or particle's ejection from the Solar system).
For a stationary system, such as the Sun, the constant number of dust grains 
(being in a balance between their production by minor bodies and eventual 
disappearance in due course of a drift toward the Sun under the P-R drag), 
and no planets included, this approach would not 
require a detailed proof. For a system that incorporates several 
planets, any  particular grain's trajectory cannot be considered as a 
stationary one, the cloning of one dust particle     
into many others as outlined above is justified by the following arguments.

After a dust particle starts its journey,
gravitational scattering on the planets causes {\it chaotic} change
of the particle's orbital elements. For example, we found that the orbital 
parameters of a particle of 
$r=5$-$10\mu$m changed by $\gax 1$\% in $10^3$ yrs and by  $\gax 10$\% in 
$10^5$ yrs. In other words, during the particle's lifetime of $2\cdot 10^7$
yrs, there are hundreds of strong changes in its trajectory. Due just to
gravitational scatterings on the giant planets, the scattered particles 
rapidly forget their initial conditions (Levison \& Duncan 1997). 
Unless the number of particle trajectories is very small,
the computed distribution function of dust particles depends only weakly
on initial $(a,e,i)$-orbital elements for the particular trajectory,
and practically does not depend on the time the trajectory starts.
Therefore,  in deriving the particle distribution function for such a highly 
chaotic system as the dust population in the outer Solar system,  
 the initial positions of kuiperoids as the dust sources are much more
important than the start time for particles.

For the present paper, we simulated 280 trajectories of dust particles
starting from the apocenter and pericenter of 100 KBOs, which
produced $1.2\times 10^{11}$ particle positions.
In the course of our computations of each particle's trajectory, the 
following procedure was applied:
\begin{enumerate} 
\item {} 
The computed particle positions were sorted on a 3D-grid containing
$45\times 180 \times 244 \approx 2\cdot 10^{6}$ cells with  steps in
(heliocentric latitude $\varphi$, longitude $\lambda$, and
radius $R$) of ($2^\circ,~2^\circ,~0.025~R$ [AU]).

\item {} 
The computer recorded the particle's $(a,e,i)$ orbital elements and
$(x,y,z)$-coordinates once per  revolution of Neptune
around the Sun (i.e. every 164.8 years).
This enabled us to create an auxillary file containing $1.2\cdot 10^{11}/
6\cdot 10^3=2\cdot 10^7$ 
particle positions. These coordinates  were sorted 
into two 2D data files: a $100\times 1000$ array in the $(a,e)$-plane
($a <150$ AU) and a $180\times 1000$ array in the $(a,i)$-space.
The following bins were used: $\Delta a = 0.15$ AU, $\Delta e =0.01,
~\Delta i=0.5^\circ$.
A few auxiliary 1D files dealing with particle
distributions in semimajor axis, $n(a)$, in
pericentric distance, $n(q)$, and in radius, $n(r)$, all used
$\Delta = 0.3$ AU.
\end{enumerate} 
		    
Let us estimate the number of trajectories which would be {\it sufficient}
to derive the stationary distribution function. Suppose that
every cell of the  $(a,e,i)$-space or  $(x,y,z)$-space
is permeated by $N$ trajectories. The {\it necessary} condition to get 
a robust distribution of particles in any cell is that a very large
 number of trajectories, $N>>1$, visited that cell. For practical reasons,
we adopt $N\sim 10$.
Since the trajectories are  highly chaotic, it does not matter
whether a particular cell is visited, say, 10 times by the same particle or
by 10 different particles just once. We found that
a particle of $r=5$-$10\mu$m requires about $2\cdot 10^3$ yrs 
(on average) to change one  $(a,e)$-space cell for another, so that it visits 
$\sim 10^4$ cells during its lifetime. 
Therefore, on a 2D $(a,e)$-grid containing $100\times 1000=10^5$ cells,
one needs to simulate $\sim 100$ different
trajectories to achieve, on average, $N\sim 10$ particle visits 
per cell (actually, this number may be as large as a few hundred
in the densest regions and $\ll 10$ in the rarefied ones). 
For the present paper, we similated 
80 trajectories of $r=5$-$10\mu$m particles and 200 trajectories of 
$r=5$-$10\mu$m particles, which provides a resonable statictics with a rather
low level of noise.
     
We employ a second-order implicit numerical integrator
described in Taidakova (1997) and  Taidakova \& Gorkavyi (1999).
Its basic  features are outlined in the Appendix. As shown there, for dust
particles produced by typical kuiperoids the dissipative  integral of motion 
is conserved  with  an accuracy of $10^{-4}-10^{-3}$,
which is appropriate to explore such a highly chaotic system as dust
population in the Solar system.

\section*{5. The Results: Dust Belts and Their Resonant Structure}

Using the above computational approach, we have simulated the spatial
structure of the IPD cloud between
$\sim 0.5$ and $\sim 100$ AU, and determined distributions 
in the phase space of orbital elements. Here we present the results for
 dust particles of radius 1-2 $\mu$m produced by 100 
Kuiper belt objects, and  particles of radius 5-10 $\mu$m produced 
by 40 KBOs. The initial conditions for orbit integration 
 were taken in accordance with equations (1)-(2).

The distributions  have been computed with two different
values of the P-R parameter (the radiation
pressure to gravitational force ratio) $\beta=0.057$ and 0.285, and with
the solar wind drag to the PR-drag ratio $=$0.35 (Gustafson 1994).
Details of computational runs are given in Table~1. The total CPU time
of those computations with a 450~MHz PC was about 2 months.

\subsection*{5.1. Distribution of Dust in Phase Space}

{\it Distribution of Dust in $(a,e)$- and $(a,i)$-space}.
A representative dynamical trajectory of a dust particle on the phase plane
is shown in Fig.~1, where the changing role of different forces
at different parts of the trajectory is clearly seen. The trajectory
is ended with the ejection of the particle  from the Solar system
by Jupiter, which is typical for most particles. The rest of
particles eventually find their death on the the Sun   or  the planets.
Our computations indicate that about 11.5\%   of $1$-2$\mu$m
 particles  penetrate the innermost
zone of the Solar system, whereas 88.5\% of particles are ejected.
Similarly, 13.8\%  of $5$-10$\mu$m particles
penetrate the Earth zone, whereas  85\%
are ejected from the Solar system (and
one of the 80 particles, i.e. 1.2\%, fell onto Neptune).
This $\sim 12$\%  fraction of kuiperoidal particles penetrating 
the inner Solar system is consistent with what was found by
Liou, Zook, \& Dermott (1997). This fraction is not very sensitive
to the particle size, which can be explained as followss: although smaller
particles drift more quickly toward the Sun under
the P-R drag, they are more easily ejected from the Solar
system due to the combined action of the solar radiative pressure and 
gravitational scattering by the planets.

The overall picture of dust distribution in the
Solar system obtained by summation of the computed trajectories
is shown in Figs.~2a,b  using
80 trajectories of large ($r=5$-10$\mu$m) particles
and in Figs.~2c,d using 200 trajectories of small  ($r=1$-2$\mu$m) particles.
We find the simulated dust distribution be highly non-uniform.
All three dynamical classes of dust particles listed at the end of Sect.~3
are clearly seen on the $(a,e$)-plane of Fig.~2c:
\begin{description}
\item {} -- drift particles with a maximum density at $a>45$ AU 
for $1$-2$\mu$m particles and at $a>30$ AU for $5$-10$\mu$m particles;

\item {} -- resonant particles producing numerous dense populations in the 
Neptunian and other resonances; the resonant population of large
  particles has a bigger contrast with the background than the population
 of small particles;

\item {} - scattered particles stretched along the right
  boundary of the planet crossing zone (i.e. the particle pericenters
  are located near the planet's orbit), especially for Neptune, Saturn, and 
  Jupiter. Scattered particles tend to avoid resonant orbits thereby forming
  resonant gaps. The Neptunian region possesses the
most dense population of scattered particles.
\end{description}

The dust distribution in the $(a,e)$-space
shown in Figs.~2a,c indicates that {\bf each planet governs a dust
belt that consists of both resonant  and scattered particles}. 
If the resonant particles dominate in the belt, then the latter is
associated with a spatial excess of dust particles (like the
Neptunian belt described in Sect.~5.2). A different situation arises 
if the belt consists  mostly of scattered particles: in this case, the belt 
is characterized by a density minimum (like the Jovian belt). In both cases, 
the belt is revealed and can be seen as a maximum in the particle distribution 
in distance of pericenter, as will be shown at the end of this Section. 
A criterion for a dust particle to be included into the belt is
a substantial dynamical interaction of the particle with the host planet,
either by resonance or by gravitational scattering.

As can be seen from the dust particle distribution on the $(a,i)$-plane,
the inclinations of kuiperoidal  particles, on average,
substantially increase due to gravitational scattering
as the particles move from the Kuiper belt toward Jupiter.

{\it Distribution of Dust Particles in Semimajor Axis and
Their Resonant Structure}.
Fig.~3 shows the distribution of kuiperoidal dust in semimajor axis.
The vertical coordinate is a measure of the number of particles
within each 0.3 AU  bin per trajectory.
Fig.~3 reveals a rich resonant structure of each dust belt.
Arrows show positions of particular resonances. One can see
that large-size particles are more easily captured in (or spend more time
within) the resonances than
small-size particles. Furthermore, the smaller the particle size,
the smaller is the  contrast  of the resonant structure to the background,
which confirms a similar conclusion reached by Liou \& Zook (1999).
Large particles with $a>50$ AU are mostly scattered particles,
therefore one can see resonant gaps in their
distribution rather than resonant peaks, whereas peaks are seen
at smaller $a=35-50$ AU.
Indeed, as Fig.~3 demonstrates, the fraction of resonant particles
in the Neptunian dust belt is very large.

The following resonant features seen in Fig.~3 are worth mentioning:
\begin{description}
\item [{\it (i)}]
in the scattered dust component, the
 gaps at the resonances  5:2, 7:2, 4:1, 5:1, 6:1 etc., are pronounced;

\item [{\it (ii)}]
for captured particles,  peaks at the resonances
6:5, 4:3, 3:2, 5:3, 7:4, 9:5, 2:1, 4:1, 5:1, 6:1, etc., are pronounced;

\end{description}

Like the cometary belts simulated in OGT (2000a,b),
 the simulated dust belts indicate a complex
structure containing many families of captured resonant
particles and gaps. While resonances in the cometary population
can form in a {\it non-dissipative} way, 
the resonant capture of dust particles occurs 
{\it dissipatively}, and this process takes place
both inside and outside the crossing zone.
For $\beta\neq 0$, resonances are shifted by a factor
$(1-\beta)^{1/3}$. The larger the value of $\beta$, the larger is
the drift velocity and the smaller is the probability of a resonant capture.

 In a resonance $(j+1)/j$, while the eccentricity is close to the maximum,
$e_{\rm max}=\sqrt {0.4/(j+1)}$ (Weidenschilling \& Jackson 1993),
the particle's resonant lifetime  is expected to be long.
The resonances seen in Figs.~2a and 2c demonstrate this kind of behaviour:
 the larger is eccentricity of particles in a given resonance,  the more
abundant is their population. 
This results in a two-hump structure seen, for large-size particle
distribution, between $q\approx 25$ and 40 AU in Fig.~4 and between $R\approx 
25$ and 45 AU  in Fig.~5. The characteristic
shape of such dust structures follows from the Kepler laws (Kessler 1981,
GOMT 1997b): The inner and outer edges of each structure are
given by $a_{res}(1-e_{\rm max})$ and  $a_{res}(1+e_{\rm max})$,
respectively. For all resonances, in a good approximation, the position
of the inner edge is $\approx 0.85a_{\rm planet}$. At this position,
we expect
to find rather sharp inner edges and steps in
the dust density distribution (somewhat
depending on the particle size) for each giant planet's  dust belt.

Direct measurements of the expected dust density
maximum at $R\approx 27-29$ AU  by dust detectors on spacecraft
would be a strong confirmation of the predicted Neptunian dust belt, as well
as its resonant nature simulated in the present work.

{\it Distribution of Dust Particles in Pericenter Distance}.
 Fig.~4 demonstrates that the pericentric distances of many
 kuiperoidal particles are located close to the orbit of
each giant planet (see also Figs.~2a,c). One important difference between   
the  resonant  and the  scattered components of a belt is in the value 
of distance of pericenter, which is $q_r\approx (0.85-1.0)a_{pl}$ for the 
resonant component (see the location of the Neptunian belt relative to
the Neptune's orbit shown as N in Fig.~4), and $q_s\approx (1.0-1.1)a_{pl}$ 
for the scattered component (see the locations of the Jovian and Saturnian 
belts relative to the respective planets  shown as J and S). 
Thus the distribution in pericentric distance provides an additional strong 
argument in favor of the four dust belts.
Earlier (OGT 2000a,b), we have found that cometary bodies 
 concentrate into  belts near each giant planet's orbit.
As  the parameter $\beta$ decreases with increase of particle size,
the  efficiency of gravitational scattering
increases,  which makes the pericentric
distribution of large dust grains   like that of comets.

\subsection*{5.2. Spatial Distribution of Kuiperoidal Dust Particles}

Fig.~5 presents the column density
of dust particles [using all recorded particle positions $(x,y,z)$]
as a function of heliocentric distance. Going inward,
the following features of dust density distribution are worth emphasizing:
\begin{description}
\item [{\it (i)}]
an approximately constant  column
density of dust (as well as volume density) between 50 and 10 AU;

\item [{\it (ii)}]
a sharp decrease of both column and volume density of dust
between 10 and 4 AU, which is due to the ejection of particles
by Jupiter and Saturn;

\item [{\it (iii)}]
an increase of column density of dust (accompanying by
an even steeper increase of volume density near the ecliptic plane) at $R<4$ AU.
\end{description}

The column density of kuiperoidal dust forms a plateau between 50 and 
10 AU, and this seems to be the most remarkable result. For 5-10~$\mu$m
dust, the value of column density is several times higher than that for
1-2~$\mu$m grains, which is explained by a slower rate of evolution of larger
size particles,
mostly due to the P-R drag. The influence of gravitational scattering and 
resonances is also different for particles of different sizes, 
so that the
shape of the  plateau, as can be seen from Fig.~5,  depends upon particle size.

The distribution of the number density of dust in the ecliptic plane, which is 
of obvious practical interest to interpret the data of
space missions, is qualitatively similar to column density graph.
This is because the particle inclination, on average, decreases outward
(see Figs.~2b,d). A region of an elevated  density
of kuiperoidal dust, which is associated with the 
 Neptune's orbits, can be seen in Fig.~5.
In the ecliptic plane, the Neptunian dust belt is expected
to have the largest number density of particles.
The major part of the simulated Neptune's dust belt consisting mostly
of resonant particles is located between
25 and 45 AU and forms a flat dense disk. The simulated Uranian, Saturnian, 
and   Jovian dust belts basically overlap and form complex
 structures which, in their central
parts, are less dense than the Neptunian dust belt.

   There is a remarkable density minimum between Mars and Jupiter.
This minimum is due to the fact that Jupiter either ejects
from the Solar system or transfers to more inclined and eccentric
orbits an appreciable part of the dust drifting toward the Sun.
An increase of dust number density going inward
 from Mars to Earth can be explained by the dominant role
of the  P-R drag here, which would result in the number density distribution
$n(R)\propto R^{-1}$ for circular orbits or $n(R)\propto R^{-(2-3)}$ for 
eccentric orbits (see GOMT 1997b and refs. therein). 

Fig.~6 shows a 2D section of  the spatial structure of kuiperoidal dust
number density
in the region up to 10 AU perpendicular to the ecliptic plane (edge-on view).
An important feature found in our simulations is 
 the existence of a new quasi-stationary, highly inclined
dust population with pericenters near Jupiter and Saturn.
This population is seen in Fig.~6 as a sharp increase in dust density
beyond the Jupiter orbit, which looks like a `Chinese wall'.
This structure is found to be  steeper and  denser for
larger particles. The quasi-stationarity of the structure
results from a balance between the tendencies for
particle's semimajor axis $a$ and eccentricity $e$ to increase due to
gravitational scattering on the planet and to decrease due to the P-R drag.
The particle inclinations  increase   substantially due
to gravitational scattering and  resonances.


\section*{6. Discussion and Conclusions}

We have explored whether the 280 particle trajectories used  in the present 
study are sufficient to provide reliable results.
As can be seen from Fig.~5, the use of just 20 particle 
trajectories already reveals the major features of large-size
particle distribution, but for increased accuracy we used 80 trajectories. 
To explore the detailed distribution of kuiperoidal dust in phase space, 
we need as many as 80
trajectories to reveal the basic resonant features (see Figs.~2a,b). The 
comparison of Figs.~2a,b with Figs.~2c,d (where 200 trajectories are used) 
indicates that a larger number of computed trajectories leads to a decrease in 
 random fluctuations; gives a clearer picture of
strong resonances free of discontinuities; reveals a larger number of weak 
resonances;  and results in a more reliable picture of resonant gaps and 
near-resonant accumulations in the scattered component of dust. Although 
even larger statistics would certainly provide further improvements, we 
do not think that it would qualitatively  change the  results.

This study employs a number of approximations and simplifications 
(circular planetary orbits with zero inclinations, neglecting
other components of dust, including the interstellar dust), which
will be treated more accurately in further work. Flynn (1996) and 
Liou et al. (1997) claim that  large ($\gax 9~\mu$m)
kuiperoidal dust is destroyed by collisions with interstellar dust.
Nevertheless, it is instructive to compare our results 
with available observational data.

 {\it Pioneers} 10 and 11 as well as {\it Voyagers} 1 and 2
detected a large number of dust
particles in the region between Jupiter and Neptune
(Humes 1980; Gurnett et al. 1997). These data show  that
the number density of dust particles in the
outer Solar system is approximately distance-independent.
As can be seen in Fig.~5, the simulated distribution of kuiperoidal dust 
can explain this constancy.
This qualitative result will be quantified in our further
work, after improving our modelling and taking into account relative 
velocities of dust particles and the spacecraft.

The approximate constancy of the dust density
between 10 and 50 AU found in the our simulations
differentiates the  kuiperoidal dust from both the
asteroidal and cometary dust, whose
number densities are known to appreciably decrease with
distance from the Sun (for interpretation, see GOMT 1997b and refs. therein).

Our simulations 
offer a high-resolution, 3-D model of the
kuiperoidal cloud on a grid  of 2 million
cells containing 115 billion computed positions of dust particles.
The major conclusion reached in the present simulations are as follows:

1. The simulated dust distribution is highly non-uniform. Moving inward,
the column and volume density of kuiperoidal dust 
is approximately constant at heliocentric distances from 50 to 10  AU,
sharply decreases between 10 and 4 AU giving a deep minimum,
after which the column density of dust increases (accompanied by
a constant or decreasing volume density
near the ecliptic plane) at $R< 4$ AU.

2. We find a new quasi-stationary, highly
inclined dust population with pericenters near all giant planets. This
quasi-stationarity results from a balance between the tendencies for the
particle semimajor axis $a$ and eccentricity $e$ to increase due to
gravitational scattering on the planet and to decrease due to the P-R drag.
The particle inclinations $i$ increase  substantially due
to gravitational perturbations from Jupiter and Saturn.

3. Most of the dust is concentrated into four belts consisting
of resonant and scattered particles associated with
the orbits of the four giant planets. These belts are chiefly of dynamical
nature. Some of them, such as the Neptunian belt, disclose first of 
all as dust structures. The others have very distinct 
distributions in the phase space: for instance, the Jovian (Saturnian)
belt is characterized by a substantial excess of highly eccentric particles 
with the maximum of dust distribution in pericentric distance at $q\approx 
5~(10)$ AU. The Saturnian belt could be discriminated by an excess of dust
particles in the resonances 4:3, 3:2, and 2:1.

4. The  simulated dust belts reveal a complex 
resonant structure containing many families of gaps and resonant maxima. 
 The particles are either dissipatively captured into exteriour 
resonances (usually outside the crossing zone) or form gaps (usually inside 
the crossing zone). 

5.  A rather long life time in each resonance, while the eccentricity is close
 to the maximal one, results in a steep rise 
of dust density at the innermost edge
of the resonant component in all dust belts, especially the Neptunian one.
Rather sharp inner edges and `steps' in the dust density
distribution are expected to characterize each giant planet's  dust belt
at 0.85$a_{\rm pl}$, where $a_{\rm pl}$ is the planet's semimajor axis.
Neptune's dust belt is expected to have both
the largest `step' and number density of particles in the ecliptic plane
(Fig.~5). Direct detection of a dust density
maximum at $R\approx 27-29$ AU in  Neptune's  zone
would test  the  simulated dust distribution.
The Neptunian and the other  belts would be a challenging
target to discover and measure by space missions.

 The resonant features of dust distributions near giant planets
 can serve as signatures of exo-planets
in the circumstellar disks (Ozernoy et al. 2000c, Gorkavyi et al. 2000a). 
The kuiperoidal dust is likely to be a contributor of the zodiacal light 
emission in the Solar system, which is analyzed in Gorkavyi et al. (2000b).

\vspace{0.1in}
{\it Acknowledgements.} This work has been supported by NASA Grant NAG5-7065
to George Mason University. N.G. acknowledges the NRC/NAS associateship.

\section*{Appendix}

To simulate the dissipationless dynamics of comets as well as dissipative 
dynamics of dust particles orbiting around a star, we use an implicit 
second-order integrator (Potter, 1973; Taidakova, 1997; Taidakova \& 
Gorkavyi, 1999; Fridman \& Gorkavyi, 1999) in a rotating (comoving with 
Neptune) coordinate system. The latter is convenient to show  the planetary 
resonant structure as a stationary pattern.
      
The equations of motion of a dust particle in the  gravitational field of 
the Sun and the planets in this coordinate 
system  take the form (Taidakova 1990, 1997):

\begin{eqnarray}
\ddot x&=&\ 2\dot y+x+F_x \nonumber \\
\ddot y&=&-2\dot x+y+F_y \nonumber \\
\ddot z&=&\quad\quad \quad \quad \quad F_z~,\nonumber
\end{eqnarray}
\nonumber \\
where
$F_x,~ F_y,~ F_z$ are the components of the sum of the gravitational forces 
and the PR-drag. Integration employs the following equations 
(Taidakova 1990, 1997):

\begin{eqnarray}
v_x^{ [n+1]}&=&{v_x^{ [n]}(1-\Delta^2 t)+
(2v_y^{ [n]}+x^{ [n+{1\over2}]}+F_x^{ [n+{1\over2}]})
\Delta t+(y^{ [n+{1\over2}]}+
F_y^{ [n+{1\over2}]})\Delta^2 t\over 1+\Delta^2 t}\nonumber \\
v_y^{ [n+1]}&=&{v_y^{ [n]}(1-\Delta^2 t)-
(2v_x^{ [n]}-y^{ [n+{1\over2}]}-F_y^{ [n+{1\over2}]})
\Delta t+(x^{ [n+{1\over2}]}+
F_x^{ [n+{1\over2}]})\Delta^2 t\over 1+\Delta^2 t}\nonumber \\
v_z^{ [n+1]}&=&v_z^{ [n]}+F_z^{ [n+{1\over2}]}\Delta t \nonumber\\
x^{ [n+1]}&=&x^{ [n]} +(v_x^{ [n+1]}+v_x^{ [n]})\Delta t/2 \nonumber \\
y^{ [n+1]}&=&y^{ [n]} +(v_y^{ [n+1]}+v_y^{ [n]})\Delta t/2 \nonumber \\
z^{ [n+1]}&=&z^{ [n]} +(v_z^{ [n+1]}+v_z^{ [n]})\Delta t/2 \quad, \nonumber
\end{eqnarray}
where $\ \ x^{ [n+{1\over2}]}=x^{ [n]}+v_x^{ [n]}\Delta t/2\ ;
\ \ y^{ [n+{1\over2}]}=y^{ [n]}+v_y^{ [n]}\Delta t/2$ \ ;
$\ z^{ [n+{1\over2}]}=z^{ [n]}+v_z^{ [n]}\Delta t/2\ ;$ and
$\ \ F_{x,\ y,\ z}^{ [n+{1\over2}]}=F\left(x^{ [n+{1\over2}]},
y^{ [n+{1\over2}]},z^{ [n+{1\over2}]},t^{ [n+{1\over2}]}\right) \ .$

We have tested our integrator for a non-conservative
system that includes the Sun and a test particle and is
governed by the Poynting-Robertson drag (two values 
of the parameter $\beta =$ 0.285 and 0.057 have been used). As initial
conditions for \{$a,e,i$\}, we used the following sets:
\{39 AU, 0.25, $10^\circ$\} and \{45 AU, 0, $10^\circ$\}.
The integration time step was taken in the range 16 to 160 days (it was 
smaller, the closer the test particle to the star).  Fig.~7 shows  
the accuracy of our integrator evaluated by the change of the
first dissipative integral of motion with time,
$\delta {C}=C(t)-C(0)$, where $C={a{(1-e^2)}\over{e^{4/5}}}$ 
(see, e.g., GOMT 1997b). It can be seen that, as the particle approaches 
the star, the integration error increases, but it never  exceeds 1\%
and is much better during the larger part of the trajectory. We 
 emphasize that, since in the real Solar system any dust
particle experiences strong gravitational perturbations from the planets
so that the particle's trajectory is highly chaotic,
an accumulation of dissipative integral errors as small as shown in Fig.~7 
is of no importance.                                  

Thus, our integrator takes into account  
close approaches with planets, which occur  frequently for dust particles
spiraling toward the star due to the PR-drag, and demonstrates good results 
in terms of stability of the integration error. In addition,
 this integrator is $1.5-1.9$ times faster than the ordinary 
2nd-order Runge-Kutta integrator (see Taidakova 1997).
Although our integrator is not as speedy as symplectic integrators,
it has an important advantage since it is applicable to dissipative systems,
along with non-dissipative ones.
     
     We found that our integrator employed in a non-rotating reference frame
has a similar accuracy but is several times faster than in a rotating 
system (Taidakova et al. 2000, in preparation). 
As an example of the use of our integrator in a non-rotating frame,
we have recently simulated a warp observed in the circumstellar disk of 
Beta Pictoris (Gorkavyi et al. 2000c).
     
\newpage
\centerline{\bf References}

\def\ref#1  {\noindent \hangindent=24.0pt \hangafter=1 {#1} \par}
\smallskip
\ref{Backman, D.E., Dasgupta, A. \& Stencel, R.E. Model of a Kuiper belt
small grain population and resulting far-infrared emission. 1995, ApJ 450, 
L35-L38}
\ref{Brown, W.R. \& Luu, J.X. 1998. Properties of model comae around Kuiper 
belt and Centaur objects. Icarus 135, 415-430}
\ref{Divine, N. Five populations of interplanetary meteoroids. 1993, J. 
Geophys. Res. 98E, 17029-17048}
\ref{Flynn, G.J. 1996, Sources of 10 micron interplanetary dust: the
contribution from the Kuiper belt. In {\it Physics, Chemistry, and Dynamics
  of Interplanetary Dust}, ed. B. Gustafson \& M. Hanner, (San
  Francisco: ASP), ASP Conf. Ser. 104, p.~171-175}
\ref{Fridman, A.M. \& Gorkavyi, N.N. 1999, {\it Physics of
   Planetary Rings. Celestial Mechanics of a Continuous Media}.
   Springer-Verlag, pp. 436}
\ref{Gorkavyi, N.N., Ozernoy, L.M. \& Mather, J.C. 1997a,
   A new approach to dynamical evolution of interplanetary dust
   due to gravitational scattering.  ApJ 474, 496-502}
\ref{Gorkavyi, N.N., Ozernoy, L.M., Mather, J.C. \& Taidakova, T.
  1997b, Quasi-stationary states of dust flows under
  Poynting-Robertson drag: new analytical and numerical
  solutions, 1997, ApJ 488, 268-276}
\ref{Gorkavyi, N.N., Ozernoy, L.M., Mather, J.C. \& Taidakova, T.
 1998a, Structure of the zodiacal cloud: new analytical
 and numerical solutions. Earth, Planets and Space, 50, 539-544}
\ref{Gorkavyi, N.N., Ozernoy, L.M., Mather, J.C. \& Taidakova, T. 1998b,
The large-scale structures in the Solar system: II. Resonant dust belts
associated with the orbits of four giant planets. {\tt astro-ph/9812480})}.
\ref{Gorkavyi, N.,  Ozernoy,   L., Mather,  J., \&
Heap, S. 2000a, ``Orbital motion of resonant clumps in dusty circumstellar
disks as a signature of an embedded planet".
WWW e-print astro-ph/0005347; In {\it Disks, Planetesimals,
and Planets"} (F.~Garzon et al., eds.)
ASP Conf. Ser. {\ul {000}}, 000-000 (in press)}
\ref{Gorkavyi, N.N., Ozernoy, L.M., Mather, J.C. \& Taidakova, T. 2000b,
 The NGST and the zodiacal light in the Solar system. In
 {\it NGST Science and Technology Exposition} (eds. E.P. Smith 
 \& K.S. Long), ASP Series, 207, 462-467 ($\equiv$ {\tt astro-ph/9910551})}
\ref{Gorkavyi, N.N., Heap, S.R., Ozernoy, L.M., Mather, J.C.
    \& Taidakova, T. 2000c, Model for the warp in the $\beta$ Pictoris
     disk (in preparation)}
\ref{Gurnett, D.A., Anser, J.A. Kurth, W.S., \& Granroth, L.J. 1997, 
Micron-sized particles detected in the outer Solar system by the Voyager 1 and 
2 plasma wave instruments. Geophys. Res. Lett. 24, 3125-3128}
\ref{Gustafson, B.A.S. 1994, Physics of zodiacal light. Ann. Rev. Earth Planet.
 Sci. 22, 553-595}
\ref{Hamilton, P.D. 1994, A comparison of Lorentz, planetary gravitational,
and satellite gravitational resonances. Icarus, 109, 221-240}
\ref{Humes, D.H. 1980, Results of Pioneer 10 and 11 meteoroid experiments:
interplanetary and near-Saturn. J. Geophys. Res., 85, 5841-5852}
\ref{Kessler, D.J. 1981, Derivation of the collision probability between
orbiting objects: the lifetimes of Jupiter's outer moons. Icarus 48, 39-48}
\ref{Kortenkamp, S.J. \& Dermott, S.F. 1998, Accretion of interplanetary
dust particles by the Earth. Icarus 135, 469-495}
\ref{Levison, H.F., Duncan M.J. 1997. From the Kuiper belt to
Jupiter-family comets: the spatial distribution of ecliptic comets.
Icarus 127, 13-32}
\ref{Liou, J.-C. \& Zook, H.A. 1997. Evolution of interplanetary dust 
particles in mean motion resonances with planets. Icarus, 128, 354-367}
\ref{Liou, J.-C.  \& Zook,  H.A. 1999, Signatures of the giant planets 
imprinted on the Edgeworth-Kuiper belt dust disk. Astron. J. 118, 580-590}
\ref{Liou, J.-C., Zook, H.A., \& Dermott, S.F. 1996, Kuiper belt grains as a 
source of interplanetary dust particles.
Icarus 124, 429-440}
\ref{Liou, J.-C., Zook, H.A. \& Jackson, A.A. 1995. Radiation pressure,
Poynting-Robertson drag, and solar wind in the restricted three-body problem.
 Icarus 116, 186-201}
\ref{Marsden, B.G. 1998, MPEC 1998-v14: Distant Minor Planets}
\ref{Ozernoy, L.M., Gorkavyi, N.N., Taidakova, T.
2000a, Large scale structures in the outer Solar system:
I.~Cometary belts with resonant features associated with the giant planets.
Mon. Not. R.A.S.  (submitted) (OGT 2000a)}
\ref{Ozernoy, L.M., Gorkavyi, N.N., Taidakova, T., 2000b,
Four cometary belts associated with the orbits of giant planets: a new
view of the outer Solar system's structure emerges from numerical
simulations. ACM conference (Cornell Univ., July 26-30, 1999).
WWW e-print: {\tt astro-ph/0001316}; Planetary Space Sci. (in press) 
(OGT 2000b)}
\ref{Ozernoy, L.M., Gorkavyi, N.N., Mather, J.C. \& Taidakova, T. 2000c,
Signatures of Exo-solar Planets in Circumstellar Dust Disks. Astrophys. J. 
Lett., July 10 issue}
\ref{Potter, D.  1973, {\it Computational Physics} (John Wiley \& Sons)}
\ref{Roques, F., Scholl, H., Sicardy, B. \& Smith, B.A. 1994, Is there
a planet around $\beta$ Pictoris~? Perturbations of a planet on a 
circumstellar dust disk. 1. The numerical model. Icarus, 108, 37-58}
\ref{Taidakova, T. 1997, A new stable method for long-time
    integration in an N-body problem. in {\it Astronomical Data Analyses,
    Software and Systems VI}, ed. G. Hunt \& H.E.~Payne, (San Francisco:
    ASP), ASP Conf. Ser. 125, p.~174}
\ref{Taidakova, T. \& Gorkavyi, N.N. 1999. New numerical
    method for non-conservative systems. {\it
    The Dynamics of Small Bodies in the Solar Systems: A Major Key to
    Solar Systems Studies}, Eds.  B.A. Steves and B.A. Roy,  Kluwer
    Academic Publishers, p.~393}
\ref{Weidenschilling, S.J. \& Jackson, A.A. 1993. Orbital resonances 
and Poynting-Robertson drag. Icarus 104, 244-254}

\newpage
\centerline{TABLE 1}
\medskip
\hrule
\medskip
\centerline{Details of Computational Runs}
\medskip
\hrule
\smallskip
\hrule
\medskip
\centerline{\vbox{
\halign{ \hfil # \hfil  & \hfil # \hfil  & \hfil # \hfil
& \hfil # \hfil  
\cr
\hfil & Particle lifetime & Number of recorded  & Number of computed   \cr 
\hfil & (in Myrs) $(^1)$    & $(a,e,i)$-elements$(^2)$ &spatial positions\cr
\hfil &       \hfil         & (in $10^{6}$)     &   (in $10^{10}$)    \cr 
\noalign{\vskip 10pt}
{\ul {Grains of $r=5$-$10\mu$m}} & & & \cr
No planets,  & & & \cr 
20 grains used &  10.1& 1.2& 0.7\cr
\noalign{\vskip 5pt}
Planets included, & & & \cr 
80 grains used & 20.3& 9.9& 5.5\cr 
\noalign{\vskip 10pt}
{\ul {Grains of $r=1$-$2\mu$m}} & & & \cr  
No planets,  & & & \cr 
20 grains used &  4.1& 0.5& 0.3\cr
\noalign{\vskip 5pt}
Planets included, & & & \cr 
200 grains used & 6.7& 8.1& 6.0\cr 
\noalign{\vskip 8pt}
}}}
\hrule
\bigskip
\begin{description}
\item
[$(^1)$] until the particle impacts the Sun, or a planet, or is ejected
from the Solar system. An average value of the lifetime is given.
A larger value of the lifetime for larger-size grains is explained by a slower
PR-drag. Presence of planets additionally increases the particle lifetime
due to captures into resonances.

\item [$(^2)$] taken with the time step $=$ 1 Neptune's revolution 
about the Sun.
\end{description}

\newpage
\centerline{\bf Figure Captions}
\vspace{0.1in}
{\bf Figure 1}. Representative trajectory of a dust particle of
of $r=5$-$10\mu$m ($\beta=0.057$) on the
planes of orbital coordinates \{$a,e$\} (panel {\bf a}) and \{$a,i$\}
(panel  {\bf b}). The trajectory presents the particle positions taken 
every $5\times 10^3$ yrs. Dashed curves show the boundaries of the crossing
zones of the four giant planets.
Diamond indicates the particle's initial position. Numbers {\it 1} to
{\it 8} mark the dominating dynamical process on the given part of the
trajectory: {\it 1} -- a `jump' from the parent body (KBO) due to the
solar pressure (see Eqs. 1 and 2); {\it 2} -- drift of particles due to
the P-R drag; {\it 3} -- resonant capture into the 3:2 resonance
with Neptune, which results in a balance between the P-R drag and
gravitational influence of the planet; {\it 4-8} - gravitational scattering 
of the particle
by giant planets, with  eventual ejection out of the system  by Jupiter.
\vspace{0.1in}

{\bf Figure 2}.
 2D density of the kuiperoidal dust on the plane of orbital coordinates,
with  bins $\Delta a = 0.15$ AU, $\Delta e =0.01,
~\Delta i=0.5^\circ$. To represent the number of particles in each cell, a
decimal-logarithm grey scale is employed, i.e. each shade differs 10-fold
from the neighboring one. Numerous resonant lines and gaps are seen.
The boundaries of the crossing zones of the four giant planets are indicated
by solid curves. Positions of the first 40 ({\bf a, b}) or 100 ({\bf c,d})
Kuiper belt objects taken from Marsden (1998) are shown by diamonds.
Numerous resonant structures are clearly seen.

{\bf a, b}. 80 dust particles of $r=5$-$10\mu$m ($\beta=0.057$) start
their journey from the pericenters and the apocenters of orbits
of 40 kuiperoids.

{\bf c, d}.  200 dust particles of radius $r=1$-$2\mu$m ($\beta=0.285$) start
their journey from the pericenters and the apocenters of orbits
of 100 kuiperoids.

\vspace{0.1in}
{\bf Figure 3}.
Distribution of kuiperoidal dust in semimajor axis, 
in terms of the number of particles per bin, $n(a)$,
averaged over each trajectory. The bin size 
$\Delta a=0.01~a_{Neptune}=0.3$ AU.
     The distributions of small ($r=1$-$2\mu$m)  and large ($r=5$-$10\mu$m)
dust particles are shown by dashed and  solid  lines, respectively.
Various resonant structures are indicated by arrows, which are heavy
for large particles and thin for small ones (J, S, and U stand for the Jovian,
Saturnian, and Uranian resonances, respectively; all other
resonances are with Neptune).
\vspace{0.1in}

{\bf Figure 4}.
Distribution of kuiperoidal dust in the distance of pericenter,
in terms of the number of particles per bin, $n(a)$,
averaged over each trajectory. The bin size $\Delta a=0.3$ AU.
The distributions of small ($r=1$-$2\mu$m) and large ($r=5$-$10\mu$m)
dust particles are shown by dashed and solid lines, respectively.
 
\vspace{0.1in}

{\bf Figure 5}.
Column density of kuiperoidal dust population, in terms of number of 
particles per 0.3 AU bin   averaged over one trajectory,
as a function of heliocentric distance (solid line).
The distributions of small ($r=1$-$2\mu$m) and large ($r=5$-$10\mu$m)
dust particles are shown by thin and heavy lines, respectively.
Dotted line shows a distribution of large particles obtained with
the use of only 25\% (i.e. 20) of the available particle trajectories, which 
indicates that the results depend rather weakly  upon the number of 
 trajectories used. Dashed and dashed-dotted lines indicate the surface
density of kuiperoidal dust computed from 20
particle trajectories in the absence of planets.
Larger-size particles, due to a slower motion,
form a denser dust population. 

\vspace{0.1in}

{\bf Figure 6}.
Density profile of a small, $r=1$-$2\mu$m, kuiperoidal dust
perpendicular to the ecliptic plane, $N(R,Z)$, at heliocentric distances 
up to 10 AU. To represent the number of particles in each cell, a
natural-logarithm grey scale is used, i.e. each shade differs $e$-fold
from the neighboring one. A remarkable density minimum between Mars and 
Jupiter is clearly seen.

\vspace{0.1in}
{\bf Figure 7}.
An average numerical error in the dissipative integral of motion
along the particle trajectory (shown as a function of
semimajor axis) for a dust particle drifting under the P-R drag toward 
the Sun. Trajectories starting from the resonant kuiperoids are marked 
with {\it 1} and {\it 3}, while those starting from the flat component of KBOs 
are marked with {\it 2} and {\it 4}. Initial orbital positions 
\{$a_0,~e_0, ~i_0$\} of the parent KBOs are as follows:

1a,1p -- \{39 AU, 0.25, $10^\circ$\}, start from apocenter
and pericenter, respectively; $\beta$=0.285.

2 -- \{45 AU, 0, $i=10^\circ$\}; $\beta$=0.285.

3a,3p -- \{39 AU, 0.25, $i=10^\circ$\}, start from apocenter
and pericenter, respectively; $\beta$=0.057.

4 -- \{45 AU, 0, $i=10^\circ$\}; $\beta$=0.057.

For an overwhelming majority of trajectories, the integration errors,
$10^{-3}-10^{-4}$, are well within the acceptable limits. 

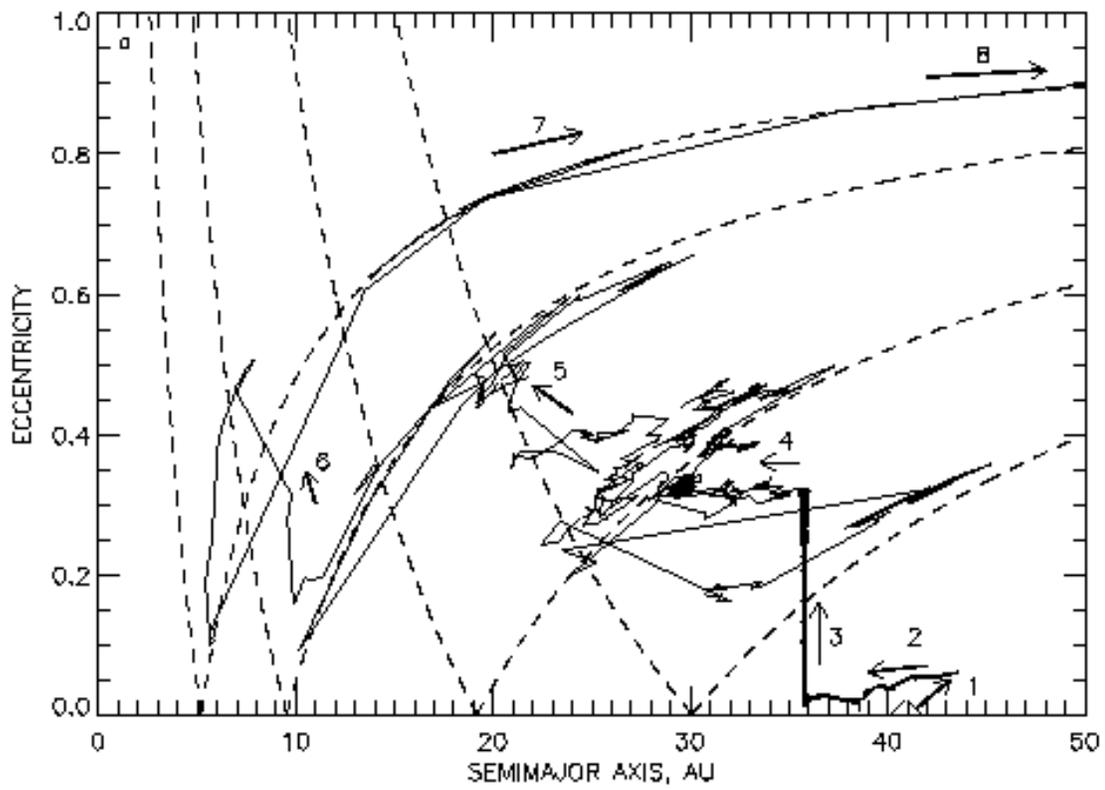
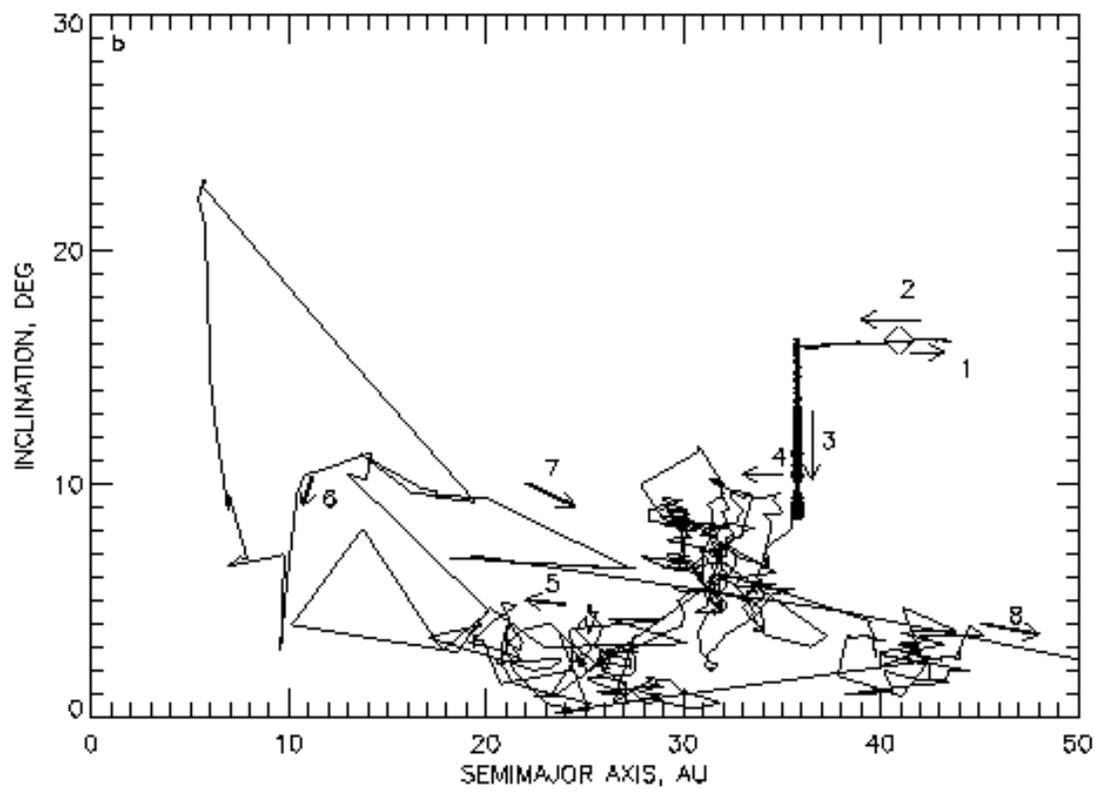
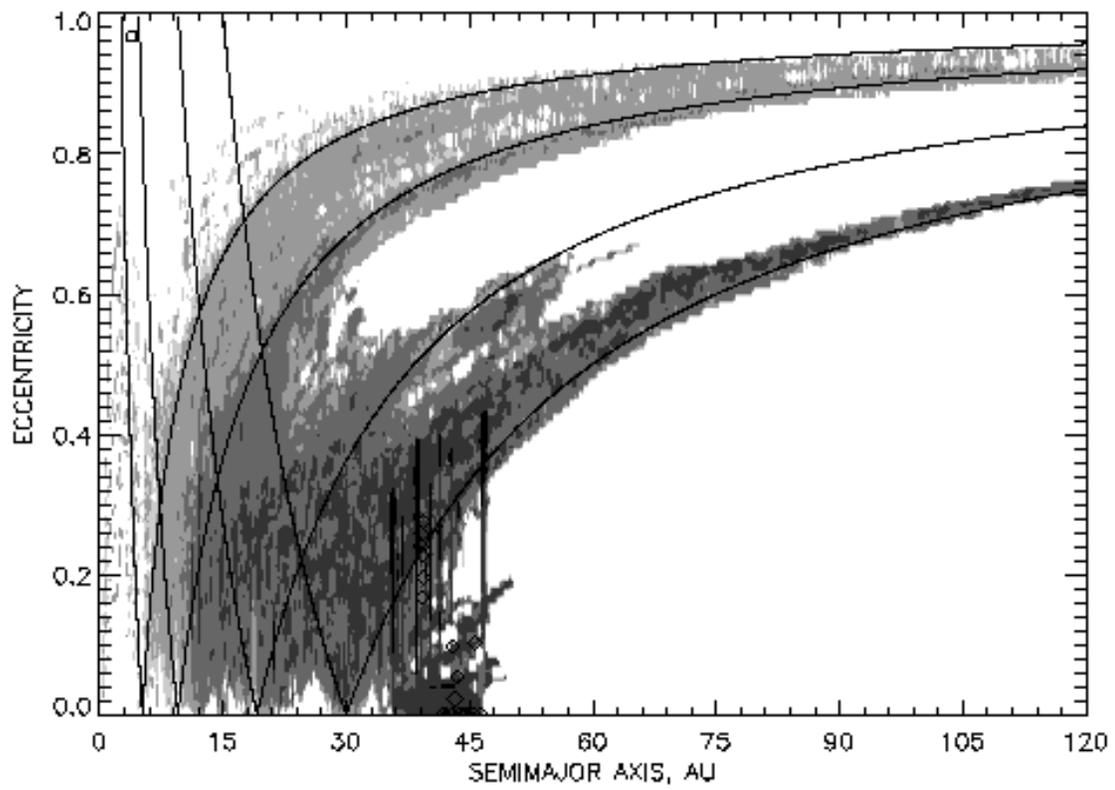
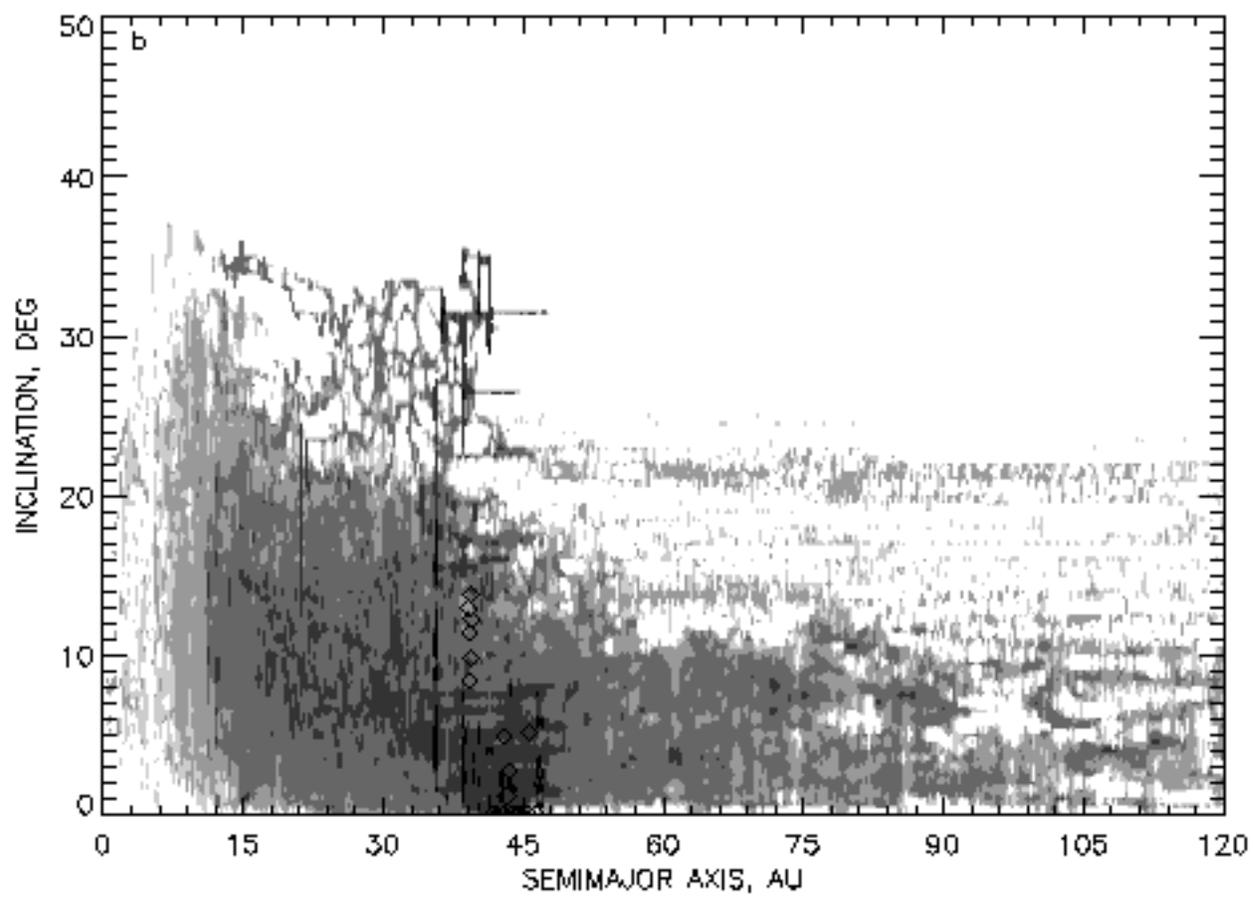
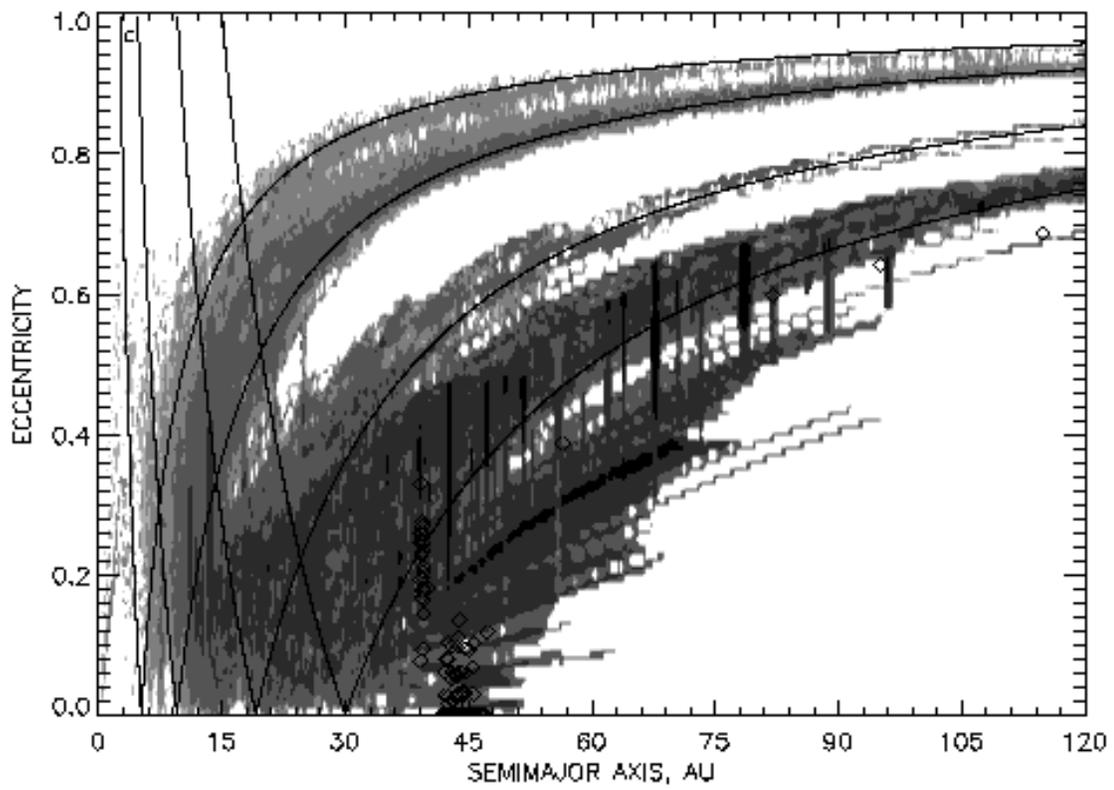
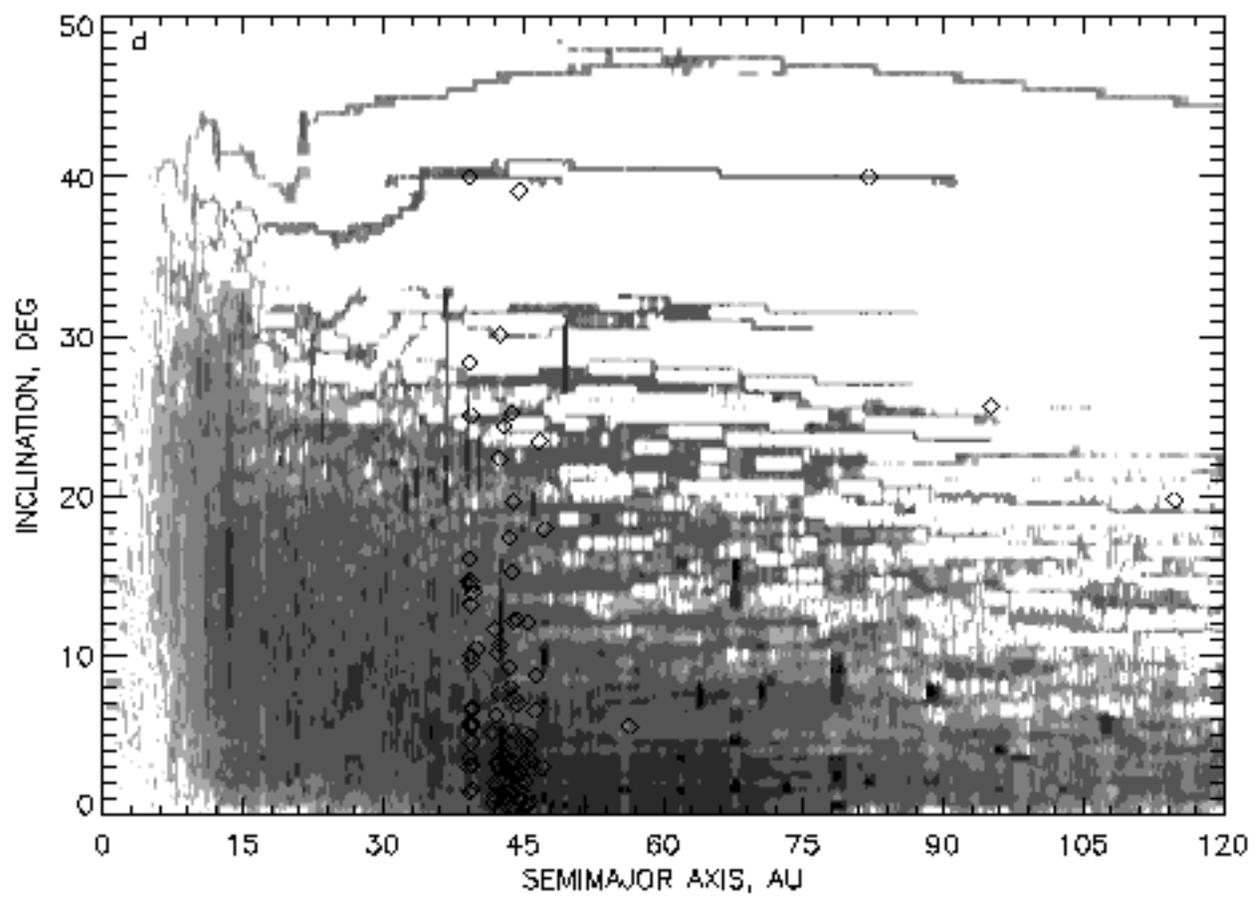
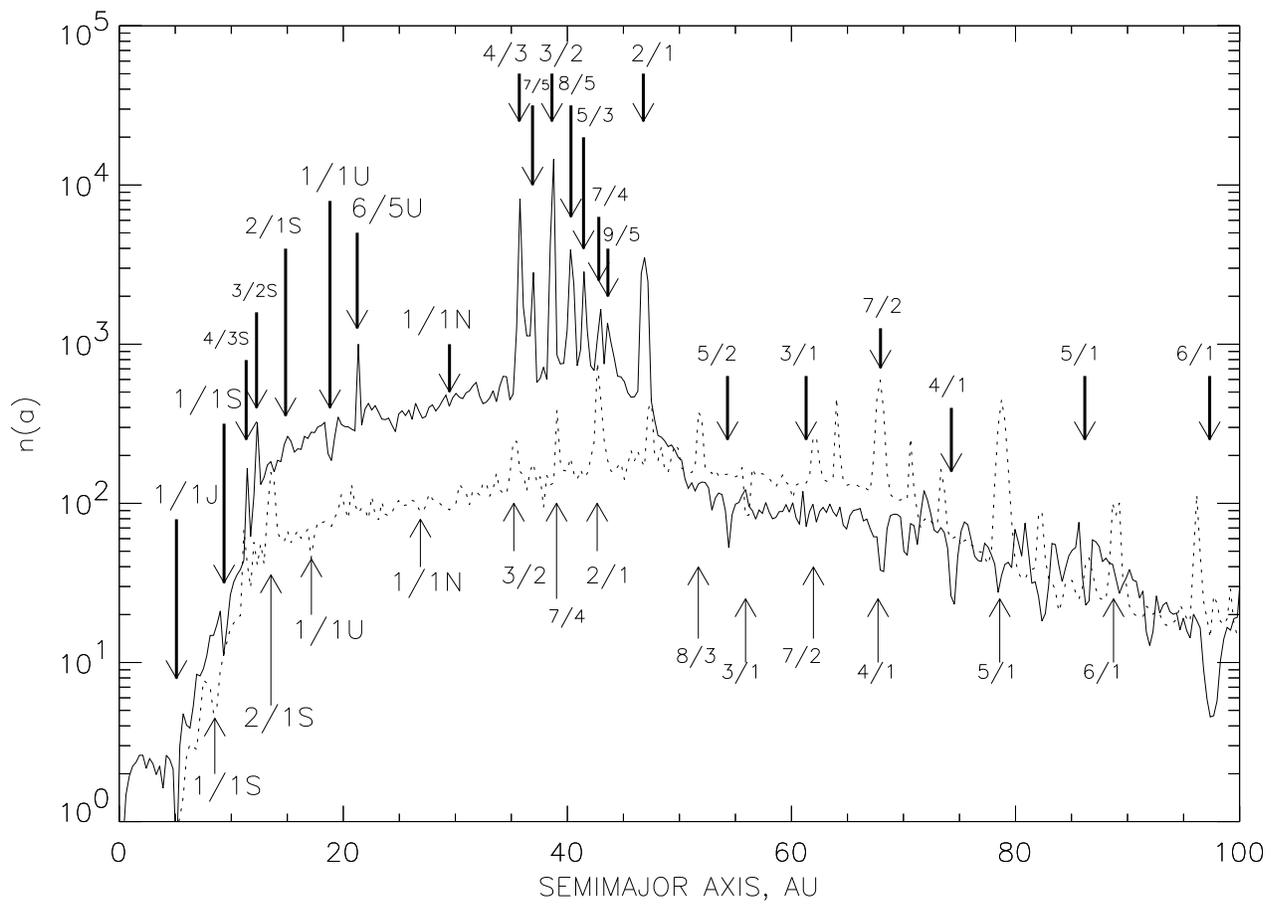
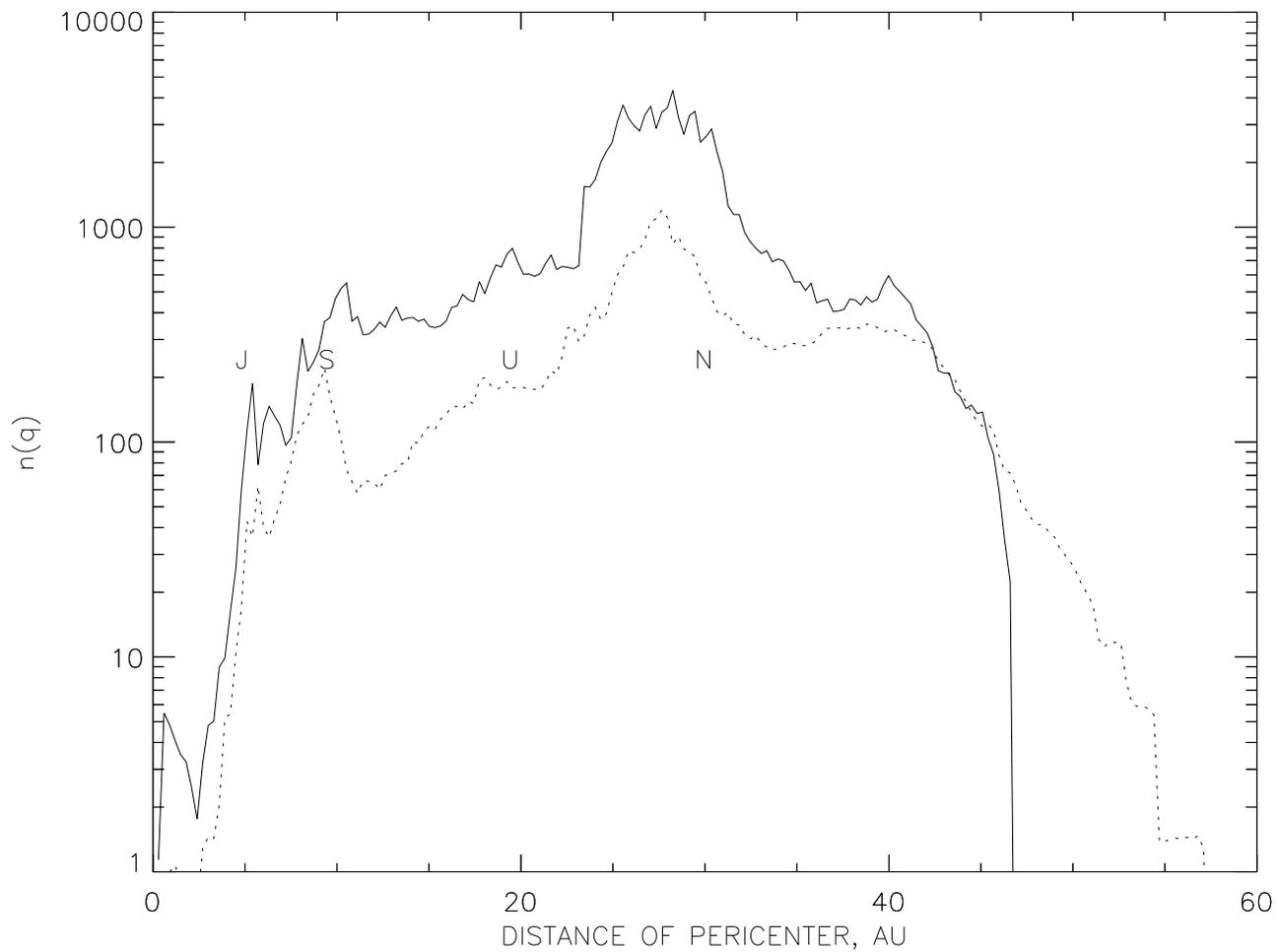
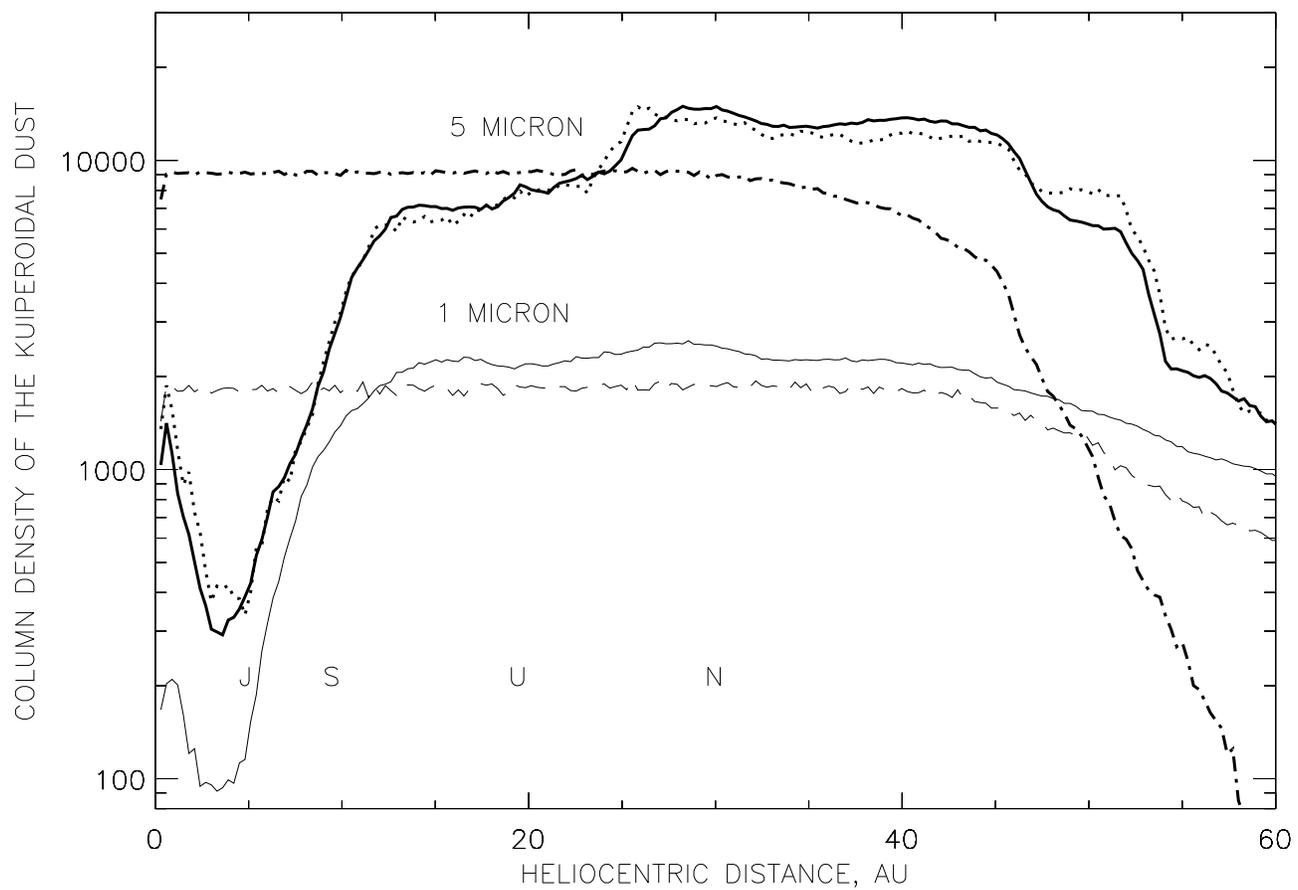
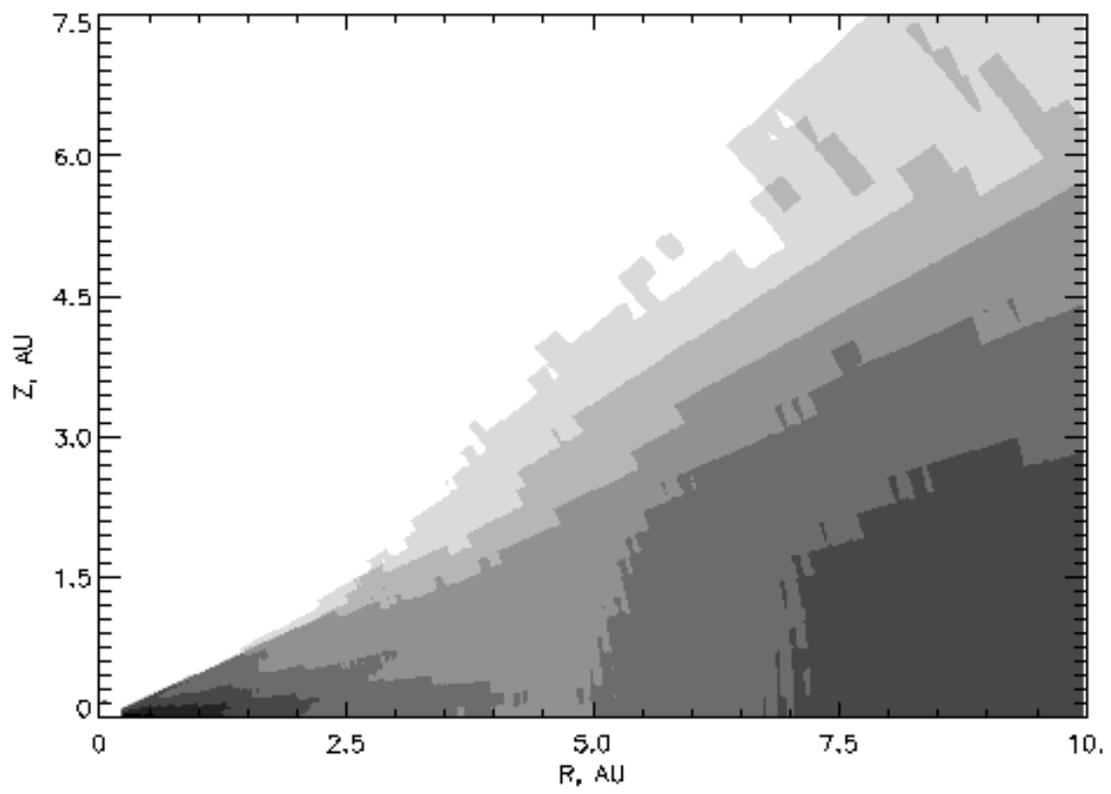
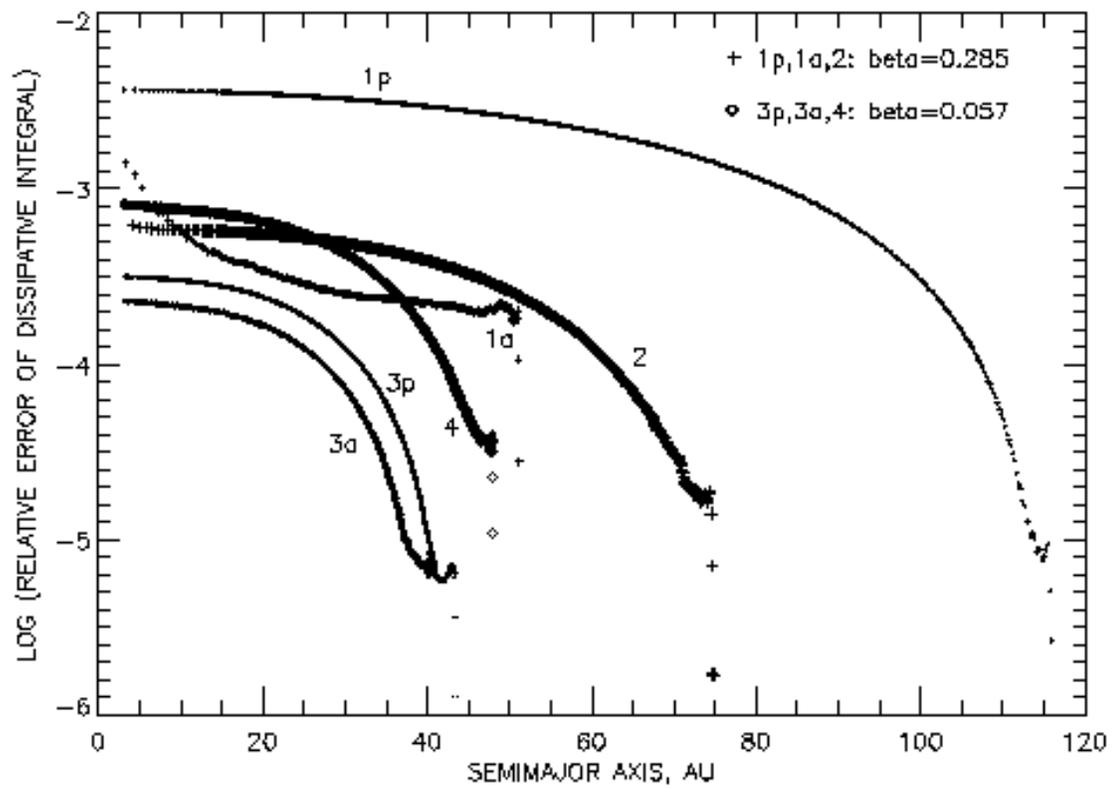
\end{document}